\documentclass[showpacs,preprintnumbers,amsmath,amssymb,12pt,floatfix,epsfig]{revtex4}
\usepackage[usenames, dvipsnames]{color}
\usepackage{amssymb}
\usepackage{bm}
\usepackage{graphicx}
\usepackage{revsymb}
\usepackage{amsmath}
\usepackage{dcolumn}

\usepackage{color} % to paint the words to be indexed

\def\be{\begin{equation}}
\def\ee{\end{equation}}
\def\bea{\begin{eqnarray}}
\def\eea{\end{eqnarray}}

\begin{document}
\draft
\title{Shape coexistence and neutron skin thickness of Pb isotopes by the deformed relativistic Hartree-Bogoliubov theory in continuum}

\author{Seonghyun Kim, Myeong-Hwan Mun and Myung-Ki Cheoun \footnote{cheoun@ssu.ac.kr}}
\address{Department of Physics and Origin of Matter and Evolution of Galaxy (OMEG) Institute, Soongsil University, Seoul 156-743, Korea}
\author{Eunja Ha}
\address{Department of General Education for Human Creativity, Hoseo University, Asan, Chungnam 336-851, Korea}

%\author{H. Sagawa \footnote{sagawa@ribf.riken.jp}}
%\address{RIKEN, Nishina Center for Accelerator-Based Science, Wako 351-0198, Japan and Center for Mathematics and Physics, University of Aizu, Aizu-Wakamatsu, Fukushima 965-8560, Japan}

\begin{abstract}
We investigate ground states properties of  Pb isotopes located between neutron and proton drip-lines estimated by two-neutron (two-proton) separation energies and Fermi energies within the   deformed relativistic Hartree-Bogoliubov theory in continuum (DRHBc). First, we report some candidates of nuclear shape coexistence in the isotope chain. They are accessed by calculating total energy as a function of the deformation parameter $\beta_2$, and for the coexistence candidates we take a couple of the deformation region bringing about minima of the energy within energy difference $\Delta E < $ 1 MeV. Second, the Pb isotopes near neutron drip-lines are also investigated and  compared to the results by other nuclear mass models. We find out eleven neutron emitters, $^{278 - 296, 300}$Pb, giving rise to the Pb peninsular near the neutron drip-line. Finally, by exploiting the neutron and proton density we deduce the neutron skin thickness (NST) of the Pb isotopes and compare to the available experimental data. The recent data regarding the shape coexistence of $^{184,186,188}$Pb and the NST of $^{208}$Pb are shown to be well matched with the present results.
\end{abstract}

\pacs{\textbf{23.40.Hc, 21.60.Jz, 26.50.+x} }
\date{\today}

\maketitle

%\section{Introduction}
Recent development of microscopic nuclear models enables us to predict many intriguing properties of unstable nuclei far from the $\beta$-stability region, such as nucleon drip-lines, shape coexistence, nuclear bubble structures and so on. In particular, the shape coexistence in heavy \cite{Pove2016,Gade2016,Heyde2011,Nach2004} and superheavy nuclei \cite{Andr2000} hints important information of nuclear shapes deeply associated with nuclear deformation and nuclear rotational band structure. The nuclear deformation is also intimately related to the surface symmetry energy in low density region which also affects the symmetry energy in the high density region like compact astrophysical objects  through the first derivative of the symmetry energy \cite{Danie2003}.

The shape coexistence is being extensively discussed through many theoretical nuclear models and experiments \cite{Heyde2011,Nach2004,Andr2000}. For example, in the nuclei having neutron number $N=$ 20 and 28, two-particle two-hole ($2p-2h$) configurations of neutrons are shown to be capable of inducing a coexistence shape of some $0^+$ states and may give rise to the inversion of island near the neutron drip-line of $N=$ 20 nuclei, and also enable us to explain the collapse of the magic number of $N=$ 28 due to the deformation in the structure of $^{43}$S \cite{Paul2009}. For heavy nuclei, such as $^{184,186}$Pb isotopes, two-quasiparticle and four-quasiparticle configuration by protons may also cause such coexistence \cite{Duguet2003,Hell2005}. Indeed, the energy spectrum of $^{186}$Pb shows such shape coexistence due to the $\beta_2$ deformation, by which most of the experimental rotational band structures of $^{186}$Pb can be understood qualitatively \cite{Duguet2003,Hell2005}.

Drip-lines of proton and neutron of heavy nuclei are also interesting because they are closely associated with the nuclear emitter of protons and neutrons far from the beta stability region. We access the separation energies of neutrons and protons of the nuclei, by which one can list the proton and/or neutron emitter of the nuclei. All of these properties are known to be sensitive on the deformation of the nuclei as discussed in the nuclear mass models \cite{Moller2016,Koura2005,Lunn2003}.

Concurrently, neutron skin thickness (NST) of $^{208}$Pb becomes one of the important data for understanding the heavy nuclear structure as well as neutron star properties. Recent data from PREX II \cite{Adhi2021,Brend2021} demonstrated more clear data of the NST of $^{208}$Pb as $R_{skin} = (0.283 \pm 0.071)$ fm, which is much more precise value rather than PREX I data \cite{Prex1,Horo2012}. It could rule out many nuclear models with the neutron star data thanks to the recent gravitational wave data and X-ray observational data. Therefore direct calculations of the NST might be a prerequisite for investigating the validity of these nuclear models before the application to the interesting topics in nuclear astrophysics.

In order to consider the above discussions we need a well-refined nuclear model which has to incorporate the deformation,  the pairing correlations and the continuum through the microscopic approach  and explain the whole nuclear masses covering nuclei near drip-lines. Along this line the deformed relativistic Hartree-Bogoliubov theory in continuum (DRHBc) was developed for deformed halo nuclei in Refs. \cite{Zhou2010,Zhou2012} and recently extended \cite{Kai2020} with point-coupling density functionals. This theory is proved to be capable for a nice description of the nuclear mass with highly predictive power \cite{Pan2021,Kaiyuan2021} and successfully applied to some nuclei \cite{Pan2019,Sun2020,In2021}, which followed the previous relativistic continuum Hartree-Bogoliubov (RCHB) approach calculated in coordinate space \cite{Meng1996,Meng98} by explicitly including the deformation in a Dirac Woods-Saxon basis \cite{Zhou2003}. In this work, we introduce briefly the basic formalism of the DRHBc theory, which was succinctly summarized in Refs. \cite{Kai2020,Zhou2012}, and focus on the results of the shape coexistence and NST of the Pb isotopes.

%Experiments

The present calculations are carried out in the following relativistic Hartree Bogoliubov theory with the
density functional PC-PK1 \cite{Zhao2010},
%\begin{equation}
\begin{equation} \label{eq:hfbeq}
\left( \begin{array}{cc} h_D - \lambda &
\Delta  \\
 - \Delta^{*} & - h_D + \lambda
  \end{array}\right)
\left( \begin{array}{c}
U_{k} \\ V_{k}  \end{array}\right)
 =
 E_{k}
\left( \begin{array}{c} U_{k} \\
V_{k} \end{array}\right),
\end{equation}
where $h_D, \lambda, E_k, (U_k,V_k)$ are the Dirac Hamiltonian, the Fermi energy and the quasiparticle energy and wave function, respectively.
The paring potential $\Delta$ is given with the pairing tensor $\kappa ({\bf r}, {\bf r}^{'})$ as follws
\begin{equation}
\Delta({\bf r}, {\bf r}^{'}) = V ({\bf r}, {\bf r}^{'}) \kappa ({\bf r}, {\bf r}^{'})
\end{equation}
with a density-dependent zero range force
\begin{equation}
V({\bf r}, {\bf r}^{'}) = {V_o \over 2} ( 1 - P_{\sigma}) \delta ( {\bf r} - {\bf r}^{'}) ( 1 -  { \rho (\bf r) \over \rho_{sat}} )~.
\end{equation}

\begin{figure}
\centering
\includegraphics[width=0.35\linewidth]{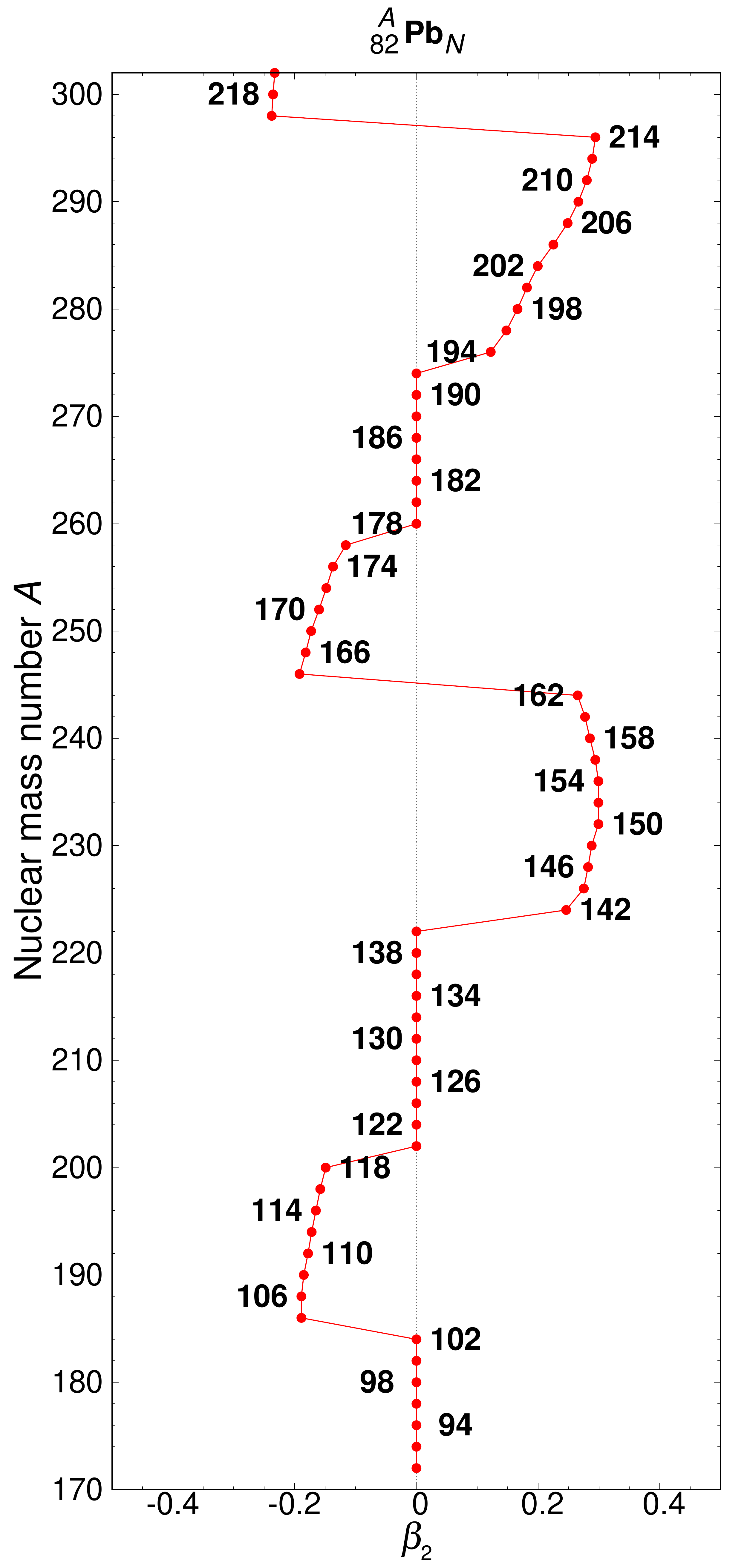}
\caption{(Color online) Deformation parameter $\beta_2$ determined by the minimum of total (ground state) binding energy calculated by DRHBc model for the nuclei from $^{172}$Pb up to $^{302}$Pb isotopes having a neutron number $N=90 \sim 220$ considered in this work. Thick letters stand for neutron number.} \label{fig1}
\end{figure}

For the pairing strength, we use $V_o$ = -- 325.0 MeV fm${^3}$. The saturation density is adopted as $\rho_{sat}$ = 0.152 fm$^{-3}$ together with a pairing window of 100
MeV. The energy cutoff $E_{cut}^+ =$ 300 MeV and the angular momentum cutoff $J_{max} = (23/2) \hbar $ are taken for the Dirac
Woods-Saxon basis. The above numerical details are the same as those suggested in Ref. \cite{Kaiyuan2021} for the DRHBc mass table calculation.
For the present Pb isotopes, the Legendre expansion truncation is chosen as ${\lambda}_{max}$ = 8 \cite{Kai2020}.

In Fig. \ref{fig1}, we illustrate the $\beta_2$ deformation of $^{172-302}$Pb isotopes obtained by the minimum of the total binding energy calculated from the DRHBc framework.
First, we note that there are three spherical nuclei region around $N=$ 100, 126 and 184. In particular, it is remarkable that $N=$ 100 might be a candidate for sub-magic shell which is occupied up to $2f_{7/2}$ shell because the nuclei around $N=$ 100 disclose spherical shapes similarly to those nuclei in the vicinity of $N=$ 126 and 184. The neutron-deficient Pb isotopes show oblate deformations after the spherical shapes near $N \sim$ 100 region, and the isotopes near stability lines are settled down to the spherical type around $N \sim$ 126 region filled up to $3p_{1/2}$ shell. The isotopes of neutron-rich side go to prolate until $^{244}$Pb by the prolate sub-shell $N=$ 162 region in Nilsson diagram and move to the oblate region by the oblate sub-shell $N=$ 172 and 178 \cite{Cwiok2005}. After that the Pb isotopes go back to spherical region owing to the magic number $N=$ 184 region and  the isotopes near neutron drip-line settle down to the prolate shape, apart from $^{298,300,302}$Pb.
%But the neutron-deficient isotopes of Rn, Ra and Th in Fig.2 are prolate and go back to spherical shape around the stability line. After that, the isotopes show zigzag patterns from prolate to oblate deformation. Interestingly, the isotopes of Rn, Ra and Th on the neutron drip lines in Fig. 2 show oblate deformation in contrast to the results in Fig. 1.

\begin{figure}
\centering
\includegraphics[width=0.80\linewidth]{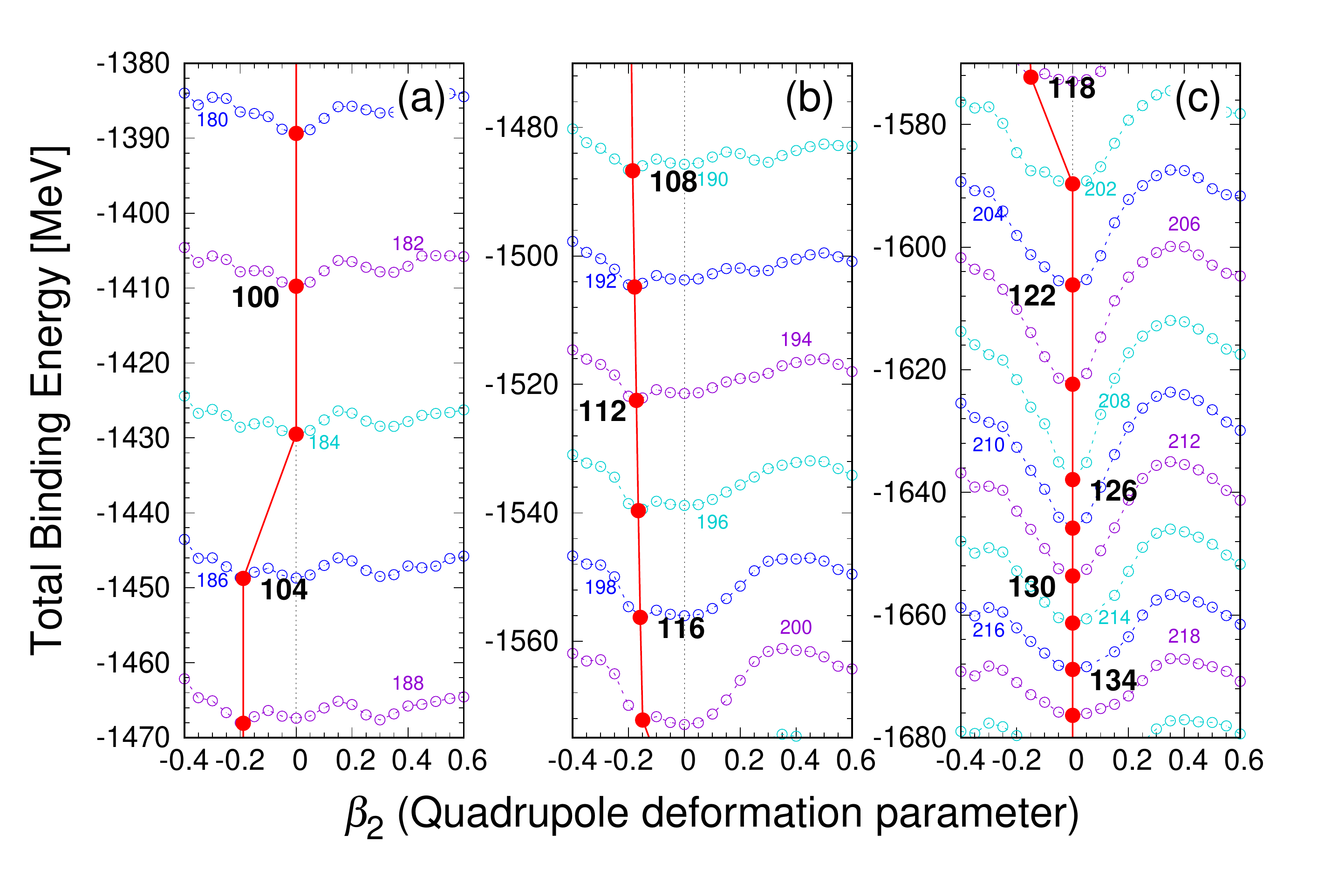}
\caption{(Color online) Evolution of total binding energy of $^{180-188}$Pb (a), $^{190-200}$Pb (b) and $^{202-218}$Pb isotopes (c) in terms of the $\beta_2$ deformation. Thick (thin) letters stand for neutron (total mass) number.} \label{fig2}
\end{figure}
%

%
%\begin{figure}
%\centering
%\includegraphics[width=0.35\linewidth]{86RN_0}
%\includegraphics[width=0.35\linewidth]{88RA_0}
%\includegraphics[width=0.35\linewidth]{90TH_0}
%\caption{(Color online)Energy surfaces of $^{86}$Rn and $^{88}$Ra and $^{90}$Th isotopes considered in this work in terms of the $\beta_2$ deformation.}\label{fig2}
%\end{figure}

In Fig. \ref{fig2}, we detail total binding energy evolution in some specific mass region $180 \le A \le 218$ in terms of the deformation parameter $\beta_2$, which shows the shape transition from spherical to oblate deformation. In the $186 \le A  \le 200$ region, the oblate shape appears. This oblate shapes are thought to come from the valence neutrons and holes between $N=$ 100 (fulfilled up to $1h_{9/2}$ and $2f_{7/2}$ shell) and $N=$ 126 (up to $2g_{9/2}$ shell). This fact is discussed with the evolution of single-particle-states (SPS) for Pb isotopes later on. Here we note that the present DRHBc model was applicable for nuclear mass table \cite{Kai2020} and the mass table of the even-even nuclei is in preparation for publication. In particular,
$^{184,186}$Pb were claimed to show a typical shape coexistence \cite{Andr2000,Duguet2003}. The total binding energy curve for $^{184,186}$Pb in Fig. \ref{fig2}(a) also displays the possibility of the shape coexistence.

\begin{figure}
\centering
\includegraphics[width=0.80\linewidth]{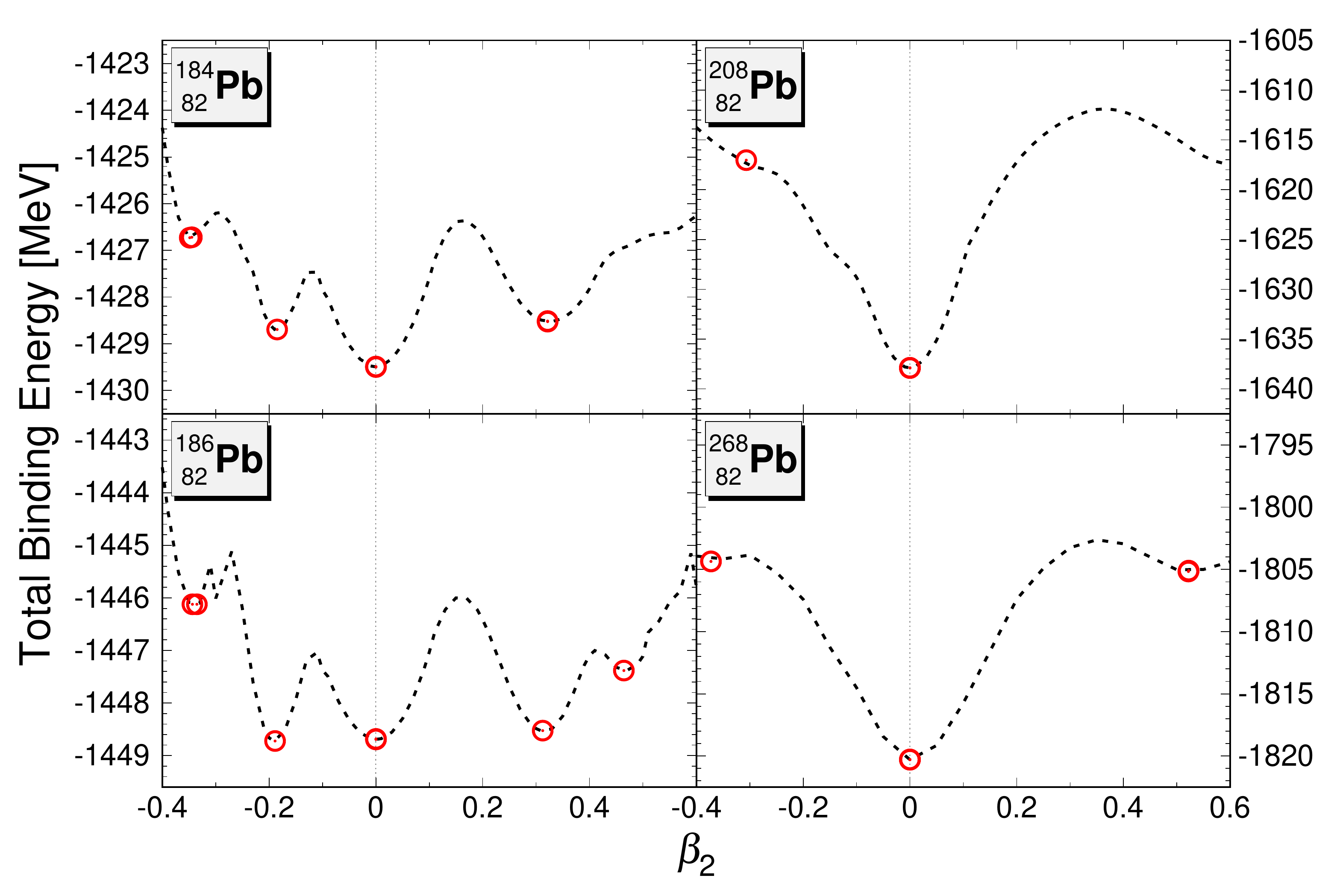}
\caption{(Color online) Detailed energy evolutions of $^{184,186,208,268}$Pb isotopes for given $\beta_2$ deformations, which show a couple of deformation region having almost same total binding energy minima for $^{184,186}$Pb. They are compared to the spherical double magic nuclei $^{208,268}$Pb. Not only global minima but also local minima points have been denoted by red circles.} \label{fig3}
\end{figure}

In Fig. \ref{fig3}, we detail the total binding energy surfaces of $^{184,186}$Pb, whose nuclei have been being discussed for their coexistence nuclear shapes \cite{Duguet2003} and those of $^{208,268}$Pb for justifying the present model for the well-known spherical nuclei. As shown in the left panels of Fig. \ref{fig3}, the results of $^{184,186}$Pb disclose a possibility of the shape coexistence of oblate, spherical and prolate shape depending on the $\beta_2$ deformation, whose energy differences are within 1 MeV. We note that the energy difference in the coexistence of $^{184}$Pb is a bit smaller than that of $^{186}$Pb.
The feasible coexistence of $^{184,186}$Pb by the oblate and prolate deformation as well as the spherical type could be a precursor of the yrast states of the rotational bands discussed in Ref. \cite{Duguet2003}. Right panels in Fig. \ref{fig3} show the spherical minima of the total binding energy surfaces of $^{208,268}$Pb near the magic number $N=$ 126 and 184 coming from the double magic nucleus.

\begin{figure}
\centering
\includegraphics[width=0.80\linewidth]{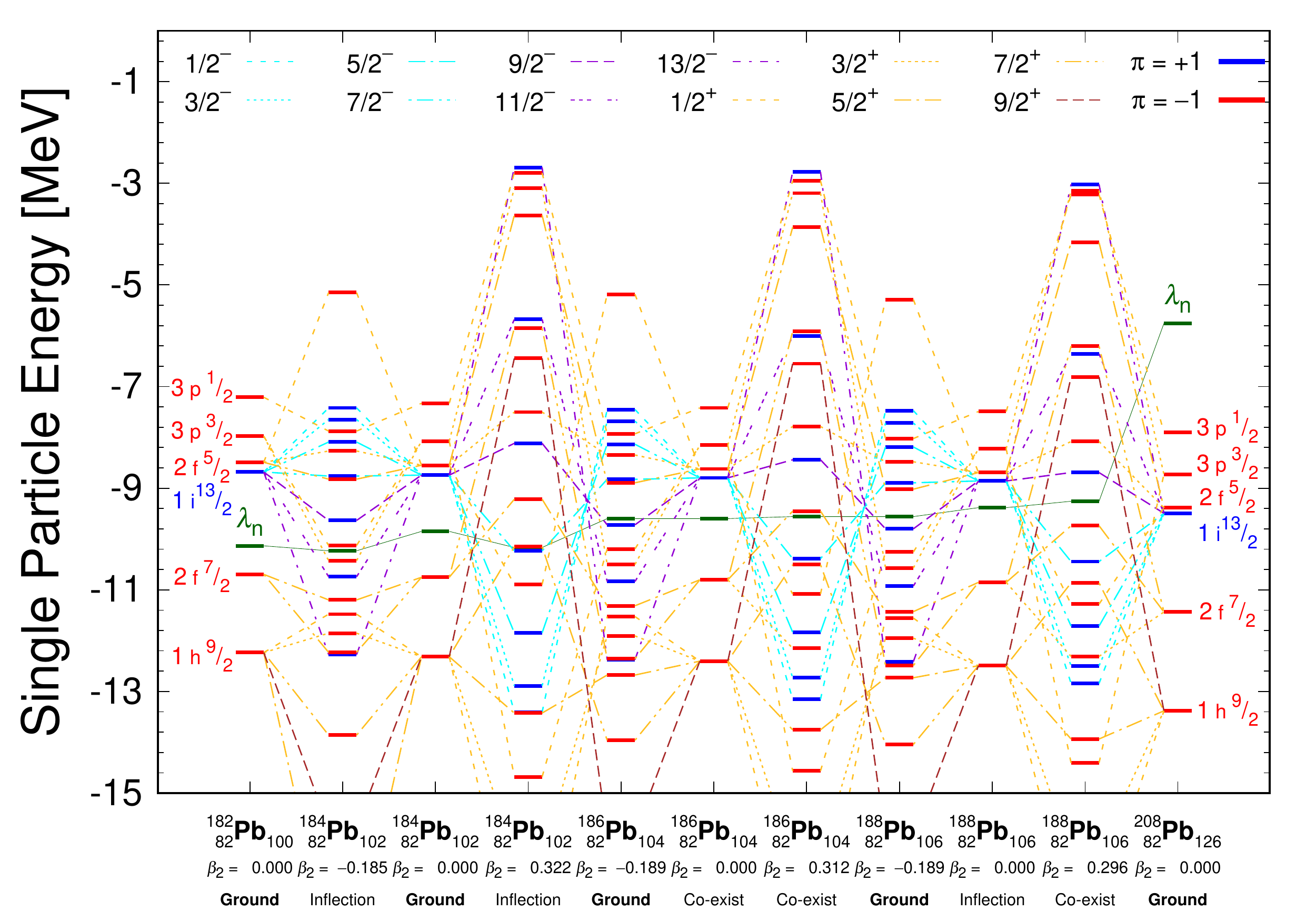}
\caption{(Color online) Neutron SPS evolutions of $^{184,186,188}$Pb isotopes for the shape coexistence : oblate, spherical and prolate case. For reference, the SPS for $^{182,208}$Pb are also shown in the leftmost and rightmost column, respectively. Fermi energies of neutron $(\lambda_n)$ are also shown for illustrating the shape coexistence.} \label{fig4}
\end{figure}
%
%%%%%%%%%%%%%%%%%%%%%%%%%%%%%%%%%%%%%%%%%%%%%%%%%%%%%%%%%%%%

The present results for the shape coexistence can be understood by the neutron SPS evolution along with the deformation in Fig. \ref{fig4}, where the SPS evolution of three isotopes $^{184,186,188}$Pb are displayed with those of the spherical $^{182,208}$Pb isotopes, which have a sub-magic $N=$ 100 and a magic $N=$ 126 number, respectively, in the leftmost and rightmost column.
One can note that the three isotopes have almost same neutron Fermi energies ($\lambda_n$) for oblate, spherical and prolate cases. In particular, we confirm $^{184,186}$Pb as the feasible coexistence nuclei because the nuclei have oblate and prolate as well as spherical minimum energies with $\Delta E < 1$ MeV. We note that $^{188}$Pb might also be the candidate of the shape coexistence as discussed by the interacting boson model \cite{Hell2005}.

\begin{figure}
\centering
\includegraphics[width=0.80\linewidth]{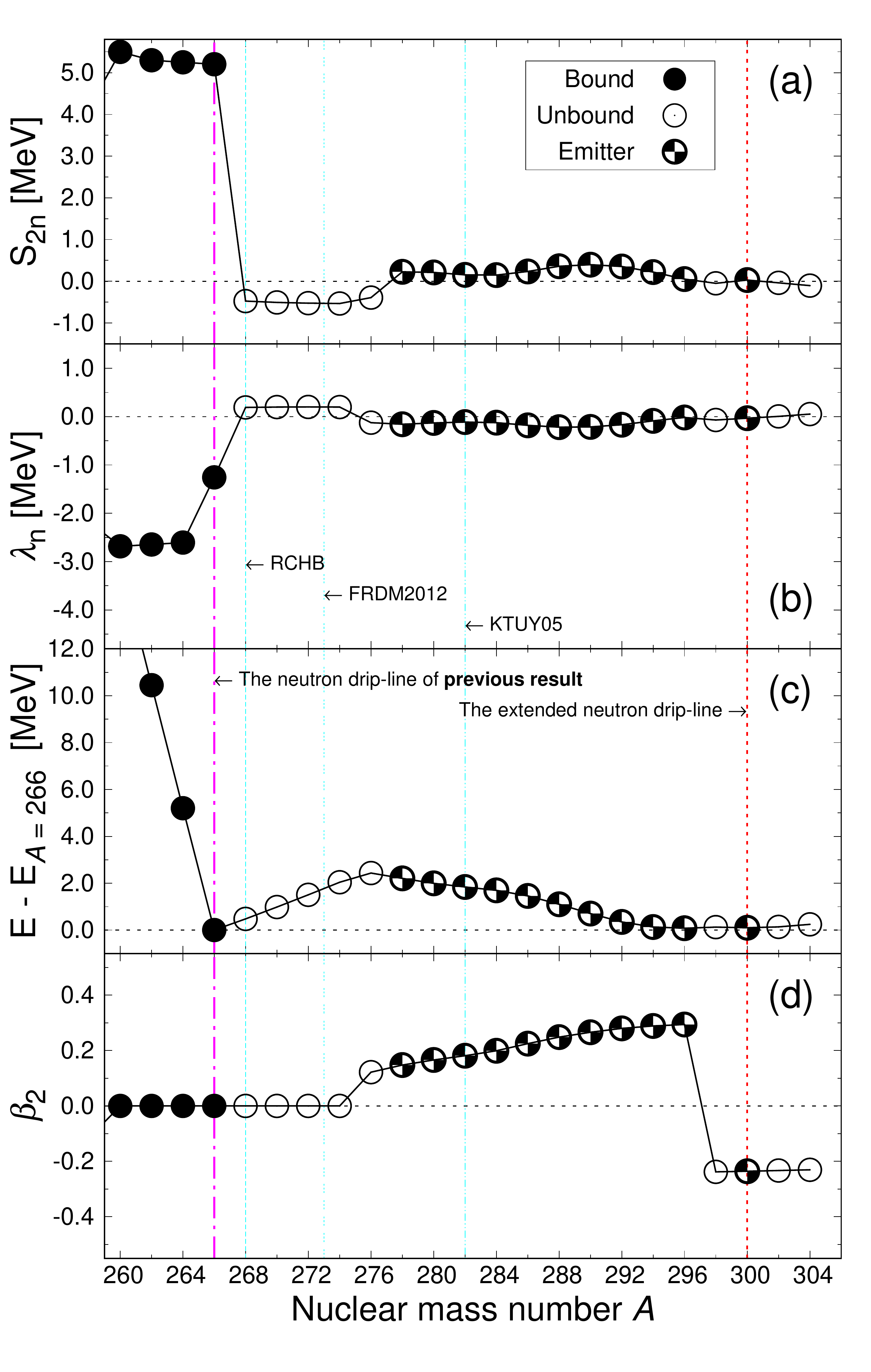}
\caption{(Color online) Two-neutron separation energies $(S_{2n})$ (a), Fermi energies $(\lambda_n)$ (b), relative energy w.r.t. the double magic nucleus, $^{266}$Pb, (c) and deformation parameter ($\beta_2$) (d) of Pb isotopes near the neutron-drip region} \label{fig5}
\end{figure}

This shape coexistence is thought to come from the deformation. The two, four and six valence neutrons in $1i_{13/2}$ shell above the core by the sub-magic number $N=$ 100 (occupied up to $1h_{9/2}$ and $2f_{7/2}$ shell) are allocated to many projection states split from the $1i_{13/2}$ shell due to the deformation.  For the prolate case, the lower ${1/2}^- \sim {7/2}^-$ (higher ${11/2}^ - \sim {13/2}^-$) projection states of $1i_{13/2}$ state are shifted to the lower (higher) energy states, as shown in the blue (purple) color system. The shifts are reversed for the oblate case. This shift of the projection states by the deformation is very general in the Nilsson diagram \cite{Nils95}. As a result, both low and high energy splitting effects by the deformation are compensated with each other and give rise to almost same Fermi energies within $\Delta E < 1$ MeV leading to the shape coexistence.

The conventional understanding of the shape coexistence is the lowering of the proton particle-hole ($p-h$) excitation for quasiparticles due to the interaction of the proton-particle with neutron-hole by the closed magic $N=$ 126 core \cite{Ione2010}. Although we do not calculate the nuclear excitation in the present work, if we treat $N=$ 100 as a kind of sub-magic shell salient in spherical case, the additional two or four neutrons above the sub-magic shell might be excited by the $2p-2h$ excitation, which may lead to the more clear shape coexistence. This coexistence near the sub-magic $N=$ 100 number might resemble the coexistence of spherical and deformed shape in $^{43}$S in the vicinity of $N=$ 28 magic number \cite{Gaud2009} leading to the collapse of the $N=$ 28 magic number \cite{Paul2009}.

\begin{figure}
\centering
\includegraphics[width=0.80\linewidth]{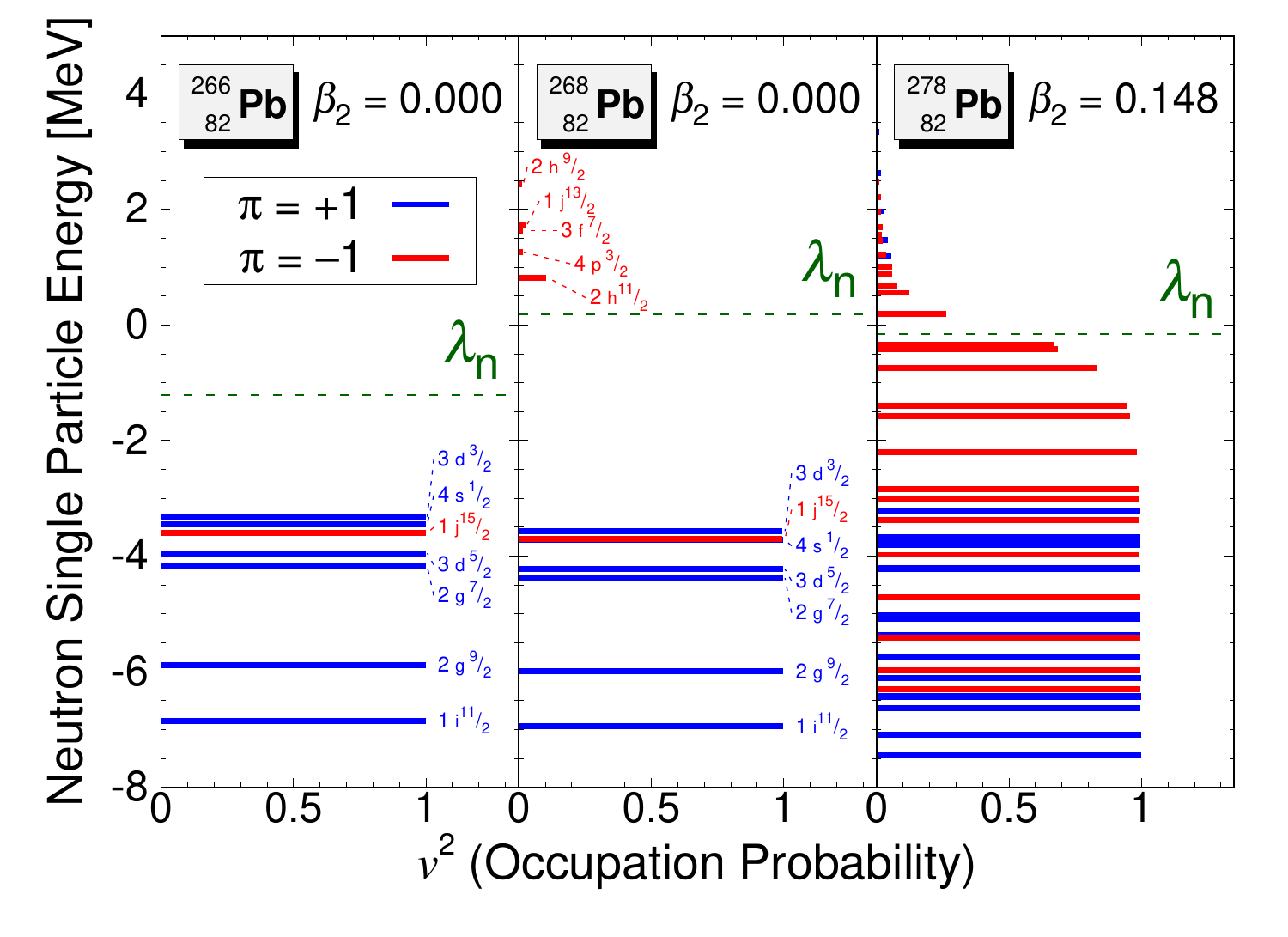}
\caption{(Color online) Occupation probabilities of bound $^{266}$Pb, unbound $^{268}$Pb and neutron emitter $^{278}$Pb} \label{fig6}
\end{figure}

Hereafter we discuss the neutron drip-line of Pb isotopes. In Fig. \ref{fig5}, we present two-neutron separation energies $(S_{2n})$ (a), Fermi energies $(\lambda_n)$ (b), relative energy w.r.t. $^{266}$Pb (c) and deformation parameter ($\beta_2$) (d) of Pb isotopes near neutron drip-line with the most neutron-rich Pb isotope estimated by other nuclear mass models, such as RCHB \cite{Meng1996,Meng98},  FRDM \cite{Moller2016} and KTUY model \cite{Koura2005}. The first point to notice is the further extension of the investigation up to $^{304}$Pb compared to those of $^{268}$Pb by RCHB, $^{272}$Pb by FRDM 2012 and $^{282}$Pb by KTUY05 model. This extension is mainly due to the deformation and the continuum states as discussed in Ref. \cite{Meng2015,In2021}.

The second point is that there exist some unstable nuclei region from $^{268}$Pb to $^{304}$Pb, which situation is more or less similar to the results of Hs near neutron drip-line as discussed in Fig.3(a) of Ref. \cite{Kaiyuan2021}, which show three unbound, five neutron emitter, {\it i.e.} stable against two-neutron emission but unstable against multi-neutron emission, with the increase of the neutron number after $^{366}$Hs having a neutron magic number $N$ = 258. This property has lead to the peninsula of heavy nuclear isotopes of $102 \le Z \le 120$ near neutron drip-lines.

\begin{figure}
\centering
\includegraphics[width=0.80\linewidth]{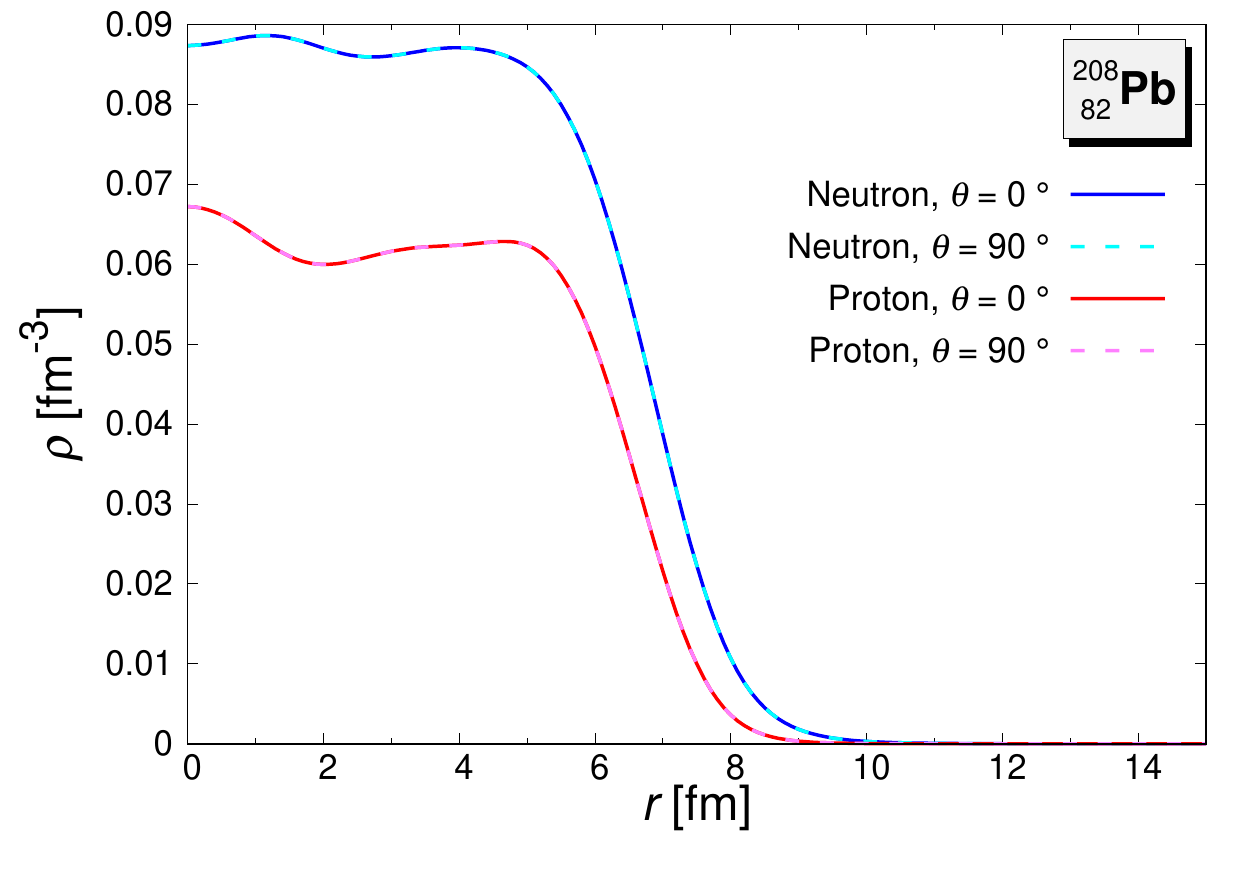}
\includegraphics[width=0.45\linewidth]{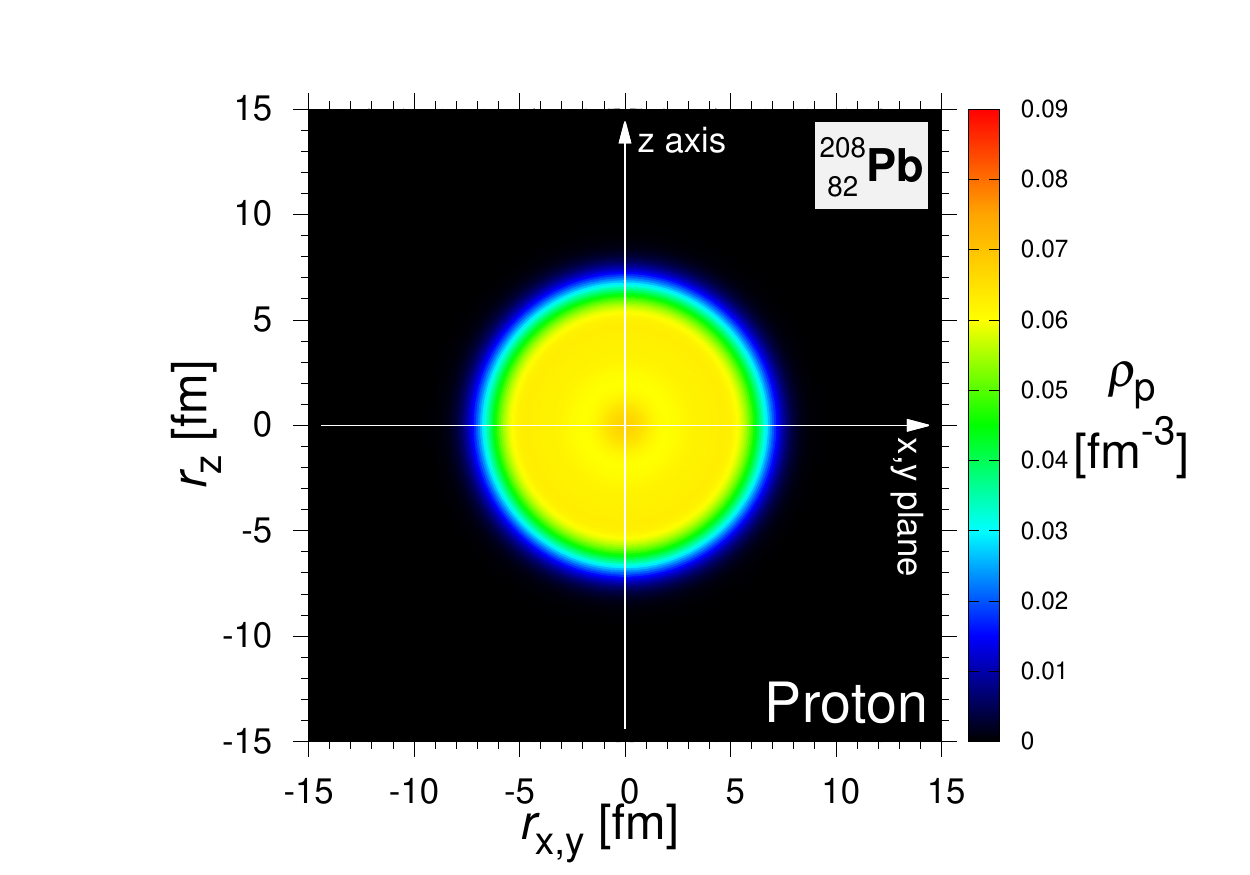}
\includegraphics[width=0.45\linewidth]{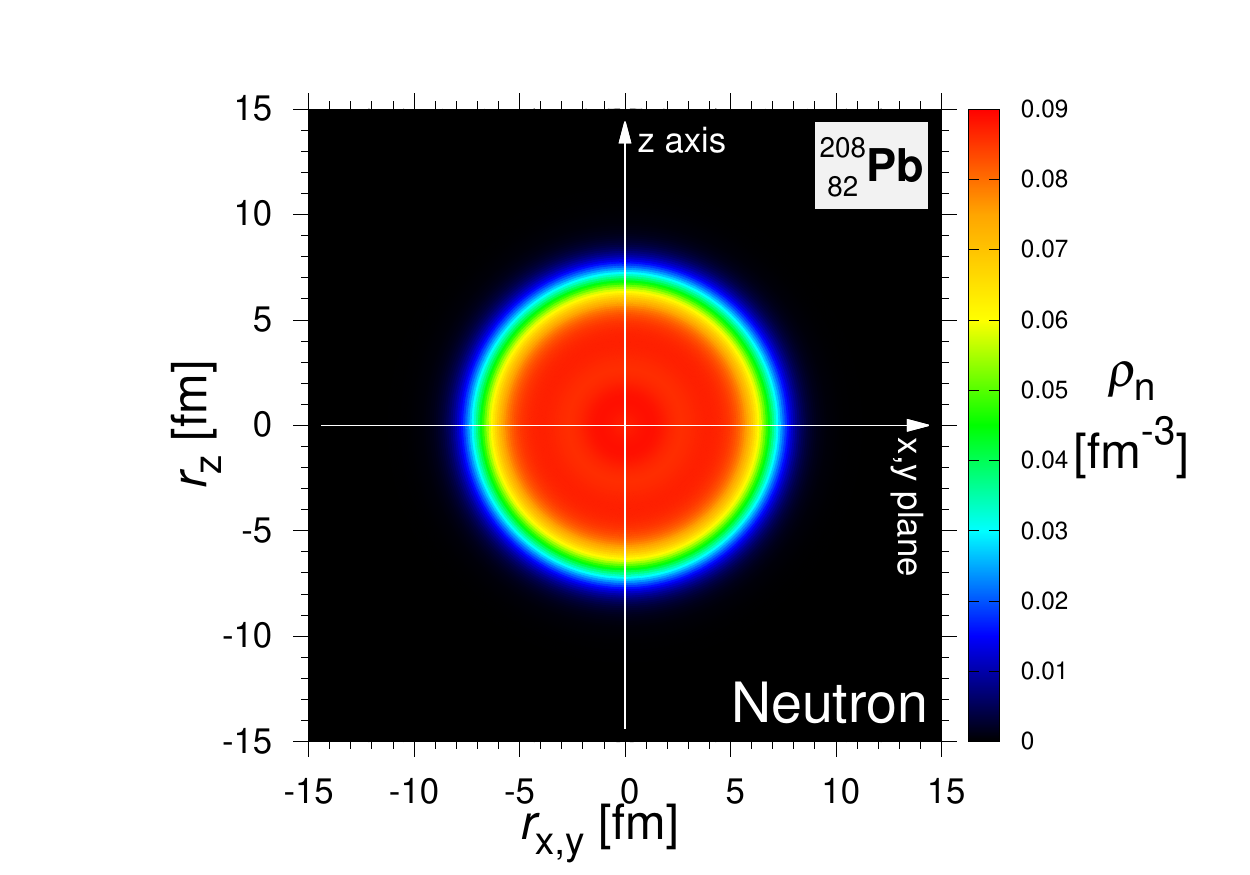}
\caption{(Color online) Upper panel is the neutron (solid line) and proton (dashed line) density distribution for the spherical $^{208}$Pb and left (right) lower panel is a bird-eye view of the neutron (proton) density distribution. Since $^{208}$Pb is a spherical nucleus, the results for $\theta =0^o$ and $90^o$ are same. }\label{fig7}
\end{figure}

In the present Pb isotopes, $^{266}$Pb having the magic number of $N=$ 184 \cite{Cwiok2005} is treated as a double magic nucleus. After $^{266}$Pb, $^{268,270,272,274,276}$Pb and $^{298,302,304}$Pb are unbound (see Fig. \ref{fig5}(a) and (b)), while other nuclei $^{278-296}$Pb and $^{300}$Pb are unstable against multi-neutron emission, as shown in Fig. \ref{fig5}(c). It means that $^{278-296,300}$Pb can be neutron emitter nuclei and become a peninsular in the nuclear chart for $Z=$ 82 isotope. This feature can be explained by the occupation probabilities of the Pb isotopes. In Fig. \ref{fig6}, the characteristics of the typical bound, unbound and neutron emitter, such as $^{266}$Pb, $^{268}$Pb and $^{278}$Pb, are clearly displayed. The neutrons beyond the double magic nucleus are shown to locate in the continuum states, which causes the unbound and neutron emitter nuclei. In particular, the large and small energy difference between the continuum states and the last occupied state leads to the unbound nuclei and the neutron emitter, respectively.

\begin{figure}
\centering
\includegraphics[width=0.80\linewidth]{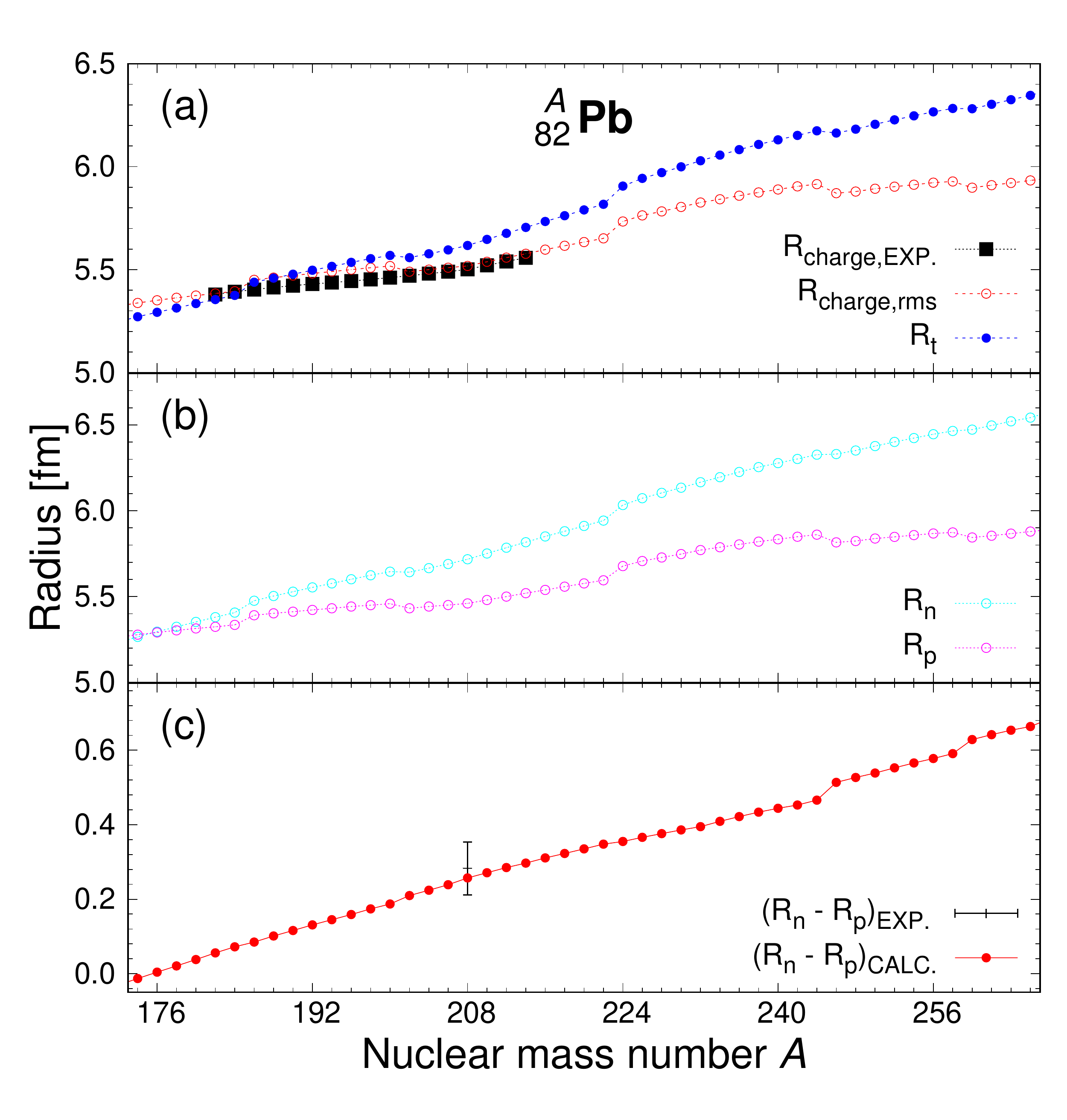}
\caption{(Color online) Evolution of charge radii calculated by the root mean square (rms) where $R_t$ is the nucleon density (a), neutron and proton radii (b) and NST (c) for Pb isotopes. Charge radii data in (a) and NST data in (c) are taken from Ref. \cite{ADNDT2013} and Ref. \cite{Adhi2021}, respectively.}\label{fig8}
\end{figure}

Finally, we show the NST of Pb isotopes. Figure \ref{fig7} illustrates the density distribution of neutrons and protons for $^{208}$Pb. Using the root mean square radius in the distribution we obtain the NST for $^{208}$Pb as $R_n - R_p$ = 0.257 fm, which is quite well matched with the PREX II experimental data 0.283 $\pm$ 0.071 \cite{Adhi2021} within the confidence level. One interesting point is that the neutron density in the inner core is much larger than that of the proton density and the surface part has also a large neutron density. It implies that the symmetry energy in finite nuclei has the contribution from both volume and surface part. The surface symmetry energy is to be subtracted from the symmetry energy  from finite nuclei for considering the symmetry energy for nuclear matter as argued in Refs. \cite{Dani2003,Niko2011}.

In addition, we illustrate the evolution of charge radii, neutron and proton density and the NST in Fig. \ref{fig8}. The charge radii obtained in the present work are quite well matched with the experimental data. Moreover, it increases with the increase of the neutron number, that is, it is swollen with the increase of the neutron number. Fig. \ref{fig8}(c) shows the NST evolution, which monotonically increases along with the neutron number. No special behavior of the NST evolution was not found. But it is remarkable that the NST of $^{246,260}$Pb show a jumping leap from the previous nuclei.
 This behavior comes from the sudden change of the nuclear shape as shown in Fig. \ref{fig1}. Also there is a kink at $A=$ 224, which may also be due to deformation effect.

In summary, we calculated the total binding energy of the Pb isotopes in terms of the $\beta_2$ deformation using DRHBc framework. Energy evolutions with the deformation were investigated in detail for those isotopes. We found some shape coexistence candidates in $^{184,186,188}$Pb, which are quite well matched the results from the rotational bands of those nuclei. We also found a peninsula in the vicinity of the neutron drip-line due to the neutron emitter  after the $N=$ 184 closed shell. They stem from the deformation and the continuum states carefully considered in the present model.  Finally, neutron skin thickness of $^{208}$Pb is calculated and compared to the recent PREX II data. The results turn out to be quite successful with the reported confidence level. Also the evolution of the NST and the related charge radii are presented for the Pb isotopes.

In conclusion, the DRHBc framework is shown to predict and properly confirm some important nuclear properties of Pb isotopes. The results of the application to other heavy nuclei will appear elsewhere and related mass table is now in preparation for publication.

\vskip1cm

Helpful discussions with members of the DRHBc
Mass Table Collaboration are gratefully appreciated. This work was supported by the National Research Foundation of Korea (Grant Nos. NRF-2018R1D1A1B05048026, NRF-2020R1A2C3006177, NRF-2020K1A3A7A09080134, NRF-2021R1F1A1060066 and NRF-2021R1A6A1A03043957). This work was supported by the National Supercomputing Center with supercomputing resources including technical support (KSC-2020-CRE-0329, KSC-2021-CRE-0126 and KSC-2021-CRE-0272).

\end{document}